\documentclass[preprint2]{aastex}

\shorttitle{Planetesimal infall in LkH$_\alpha$234}
\shortauthors{Chakraborty et al.}

\begin{document}


\title{Evidence of Planetesimal infall on to the very young Herbig Be star LkH$_\alpha$234
\footnote{Based on observations from the 9.2m Hobby-Eberly
Telescope}}


\author{Abhijit Chakraborty, Jian Ge \and Suvrath Mahadevan}
\affil{525 Davey Laboratory, Pennsylvania State University,
University Park, PA 16802} \email{abhijit@astro.psu.edu,
 jian@astro.psu.edu, suvrath@astro.psu.edu}



\begin{abstract}
We report here the first evidence for planetesimal infall onto the
very young Herbig Be star LkH$_\alpha$234. These results are based
on observations acquired over 31 days using spectroscopy of the
sodium D lines, the He I 5876\AA, and hydrogen H$_\alpha$ lines.
We find Redshifted Absorption Components (RAC) with velocities up
to 200 km/s and very mild Blueshifted Absorption Components (BEC)
up to 100 km/s in the Na I lines. No correlation is observed
between the appearance of the Na I RAC \& BEC and the H$_\alpha$
and He I line variability, which suggests that these (Na I RAC \&
BEC) are formed in a process unrelated to the circumstellar gas
accretion. We interpret the Na I RAC as evidence for an infalling
evaporating body, greater than 100 km in diameter, which is able
to survive at distances between 2.0 to 0.1 AU from the star. The
dramatic appearance of the sodium RAC and mild BEC is readily
explained by the dynamics of this infalling body making
LkH$_\alpha$234 the youngest (age $\sim$ 0.1 Myr) system with
evidence for the presence of planetesimals.
\end{abstract}





\section{Introduction}
LkH$_\alpha$234 (m$_v$=11.9mag) is a star embedded in a nebula
associated with the NGC 7129 star forming region and is at a
distance of about 1250 parsecs. This is one of the youngest known
Herbig Be objects and has an age $\sim$ 0.1 million years and
spectral type of B5 (Fuente et al. 2001). Herbig Ae/Be stars are
intermediate mass (2 to 10M$_\odot$) pre-main-sequence stars
(Herbig 1960). A number of Herbig Ae stars have been found to
possess circumstellar disks and show both photometric and
spectroscopic variability (Herbst et al. 1999). LkH$_\alpha$234
though much hotter and more massive also exhibits these
properties (Fuente et al. 2001, Herbst et al. 1999, Polomski et
al. 2002). The star shows photometric variability of 1.2 magnitude
in the V band and is known posses a circumstellar disk with the
disk mass to the stellar mass ratio (M$_D$/M$_*$) $\leq$ 0.02. A
detailed description of LkH$_\alpha$234 and its circumstellar
region can be found in Fuente et al. (2001, and references
therein).

Spectroscopic variations in metallic lines (like Na I, Ca II,
etc.) in young A type stars like $\beta$ Pictoris (20 to 100
million years) are often interpreted as the signature of star
grazing comets or falling evaporating bodies with sizes of tens of
kilometers (Grinin et al. 1996, Smith \& Terrile 1984, Thebault \&
Beust 2001). The only Herbig Be star where grazing cometary
transient events have been inferred so far is HD 100546 (Viera et
al. 1999). However, HD100546 is much older ($\geq$10 Myrs) and
cooler (spectral type $\sim$B9) compared to LkH$_\alpha$234. High
resolution spectroscopic monitoring of very young (0.1 to 1 Myrs)
Herbig Be stars are scanty and uncharted.

Here we report dramatic variations in the spectroscopic lines of
Na I 5890\AA (D2), and 5896\AA (D1), He I 5876\AA, and H$_\alpha$
of LkH$_\alpha$234 over a period of 31 days during Oct.-Nov. 2003.
The observations consists of high resolution spectra (R=30,000)
from the Hobby-Eberly Telescope (HET) using the High Resolution
Spectrograph (HRS). Section 2 describes observations and in
section 3 we discuss the results and section 4 gives the
conclusion.

\section{Observations}
We have obtained high signal to noise ratio and high resolution
spectra (R=30,000, $\sim$0.2\AA) of LkHa234 covering He I 5876\AA,
Na I D2 \& D1 lines and the Balmer H$_\alpha$ at 6563\AA. The
observations were taken with the 9.2m Hobby-Eberly Telescope (HET)
(Ramsey et al. 1998) using its High Resolution Spectrograph (HRS)
(Tull 1998). The echelle spectral orders were optimized to cover
the wavelength range between 5700 to 6750\AA. This allowed the
spectral lines of interest to be recorded by the ``blue" CCD of
the spectrograph that has fewer number of bad pixels than the
``red" CCD. The results comprise of five data sets over a time
period of 31 days. The observation dates were: 7th, 13th, 23rd,
27th October and 8th November 2003. On each night of observation,
three ten minutes exposures on LkH$_\alpha$234 were taken which
were combined to get high signal to noise ratio (S/N = 70 to 100).
All data were reduced using standard IRAF routines under ECHELLE
tasks. Every spectrum was bias subtracted, and flat-fielded.
Spectral Calibrations were done using Thorium-Argon lamp spectra
taken either immediately before or after the observations. We
found that the wavelength calibration is accurate up to 0.05\AA.

H$_\alpha$ (6563\AA), He I 5876\AA, and Na I D2 \& D1 spectral
lines are shown in figures 1 and 2 respectively and the dates of
observations are noted along side of the spectra. The x-axis is in
terms of relative velocity (km/s) with respect to the center of
the lines in the stellar rest frame and the y-axis shows line
intensities normalized with respect to the stellar continuum. In
figure 2 the zero of the relative velocity scale is with respect
to the stellar rest frame of the sodium D2 line.

\section{Results \& Discussion}
\subsection{Gaseous accretion}
We illustrate in figure 1 the variability in H$_\alpha$ and He I
lines. These are seen as stellar photospheric absorption lines in
the spectra observed on 7th Oct. 2003. However, in all the later
days of observations the H$_\alpha$ appeared to be in emission
consisting of two components, the Red shifted Emission Component
(REC), and the Blue shifted Emission Component (BEC). The
H$_\alpha$ equivalent width varied from +3\AA (on 7th Oct.) to
-50\AA (13th Oct.) in just 6 days and it continued to rise up to
-100\AA (27th Oct.) before it decayed to -74\AA (8th Nov.). The
relative velocity between the REC and BEC varied between 150 km/s
to 225 km/s with the REC stronger than the BEC by a factor of 2.5
to 3.5.

The spectra obtained on 7th Oct. 2003 do not show any sign of
emission or activity in H$_\alpha$ or He I. This observation shows
the star in a quiescent state, with neither accretion nor stellar
wind affecting the absorption line profile. The variations in the
He I line are found to correlate with the variations in the
H$_\alpha$. Whenever the H$_\alpha$ BEC and REC are stronger the
He I either fills up due to mild emission or shows a very large
absorption width (almost 1.7 times the photospheric line width).
The line width varied from 150 km/s to 260 km/s. A B5 star does
not have sufficiently energetic photons to ionize He I (Jaschek \&
Jaschek 1995), so the filling up of the He I absorption line due
to mild emission is an indicator of accretion phenomena (de Winter
et al. 1999). We find the disk luminosity to be about 1$\%$ of the
stellar luminosity (L$_*\sim$10$^3$L$_\odot$) with maximum disk
temperature of $\sim$21000K, if we consider an accretion rate of
10$^{-7}$M$_\odot$ per year (Natta et al. 2000).

Therefore, the double peak H$_\alpha$ (REC \& BEC) profiles and
the He I mild emission are because of orbiting accreting gas on to
the star and as well as outflow of circumstellar gas (BEC in
H$_\alpha$ due to accretion heat). However, when the accretion
luminosity is not strong enough to produce enough ionizing
photons, the He I line appears as a very broad absorption feature.
The broad absorption feature is the effect of the combination of
the stellar photospheric line and various unresolved circumstellar
gas components corresponding to REC and BEC of H$_\alpha$.

The effect of gaseous accretion are probably also seen on the
nebular emission lines. We found a correlation between the total
equivalent width in H$_\alpha$ and the peak nebular emission in
H$_\alpha$ and Na I. Figures 1 \& 2 show that the nebular
H$_\alpha$ and Na I lines were of minimal intensity ($\sim$1.06
above the continuum) on 7th Oct. The same lines are strongest
(H$_\alpha$ $\sim$7.9, on the blue wing of the REC, and Na I D2
line $\sim$3.0) on 27th Oct. when the H$_\alpha$ equivalent width
is maximum (-100\AA). We plan to model the impact of the episodic
accretion on the nebula using a photo-ionizing code (for eg. the
Cloudy code) which will be published along with the entire spectra
of LkH$_\alpha$234 elsewhere in the near future.

\subsection{Planetesimal infall}
Figure 2 shows the variability in the Na I D2 and D1 lines.
Redshifted Absorption Components (RAC) were observed in Na I lines
on only one night (13 Oct. 2003) with a maximum redshifted
velocity of 200 km/s. The RAC's associated with both the Na I
lines are of similar depth (for redshifts $\leq$ 100km/s) which is
an indicator of saturation (unshielded Na I column density =
10$^{12}$/cm$^2$)(de Winter et al. 1999). However, the maximum
depth of these components is about 30\% of the stellar continuum
implying partial coverage of the stellar photosphere. In addition
to Na I RAC, very mild BEC (60 to 100 km/s and 1.05 to 1.07 above
the continuum) are also seen on the 13th Oct. spectra. The
variations in He I and H$_\alpha$ are observed even when no Na I
RAC \& BEC are seen (see figures 1 \& 2). Thus the appearance and
disappearance of the Na I RAC \& BEC are uncorrelated with the
variations in H$_\alpha$ and He I line profiles. We conclude that
we witnessed a transient phenomenon on 13th Oct. 2003 whose
effects lasted a few days at most.

From the Keplerian dynamics of an infalling object (Beust et al.
2001) the most redshifted Na I absorption component velocity (200
km/s) should correspond to a distance of 0.1 AU from the star. At
this distance both the number of ionizing photons and intensity of
the stellar wind will be too high for any unshielded Na I to
survive (Sorelli et al. 1996). Their calculations using
photo-ionizing codes under local thermodynamical equilibrium (LTE)
and non-LTE show that only a solid body like a comet or asteroid
can approach a hot star this close before it disintegrates and
evaporates. Further, theoretical model calculations for stars up
to B9 show that the size of solid bodies which are able to survive
up to a distance of 0.1 AU should be at least 100 km in diameter
(Beust et al. 2001). The mild Na I BEC could then be part of the
falling evaporating body being blown away by the stellar wind or
radiation pressure. However, this will also mean that the falling
solid body may have a high eccentric orbit so that the blown away
material is not projected against the stellar surface. Our Na I
RAC and BEC observations are consistent with such an object
dissociating between 2 AU and 0.1 AU from the B5 star
LkH$_\alpha$234.

In many Herbig Ae stars, variations in Na I RACs and BAC's usually
correlates with H$_\alpha$ and He I variations (de Winter et al.
1999, Grinin et al. 1996). Such Na I variations are now thought to
be generated by magneto-hydrodynamic funnelling of neutral gas on
to the star rather than solid body infall (Beust et al. 2001, Mora
et al. 2002, Natta et al. 2000). No such correlation is present in
the current LkH$_\alpha$234 data set. Of particular significance
is the observation on 27th Oct. 2003 when the REC and BEC of
H$_\alpha$ are the strongest. The absence of any Na I RAC and BEC
in this observation is a strong argument against the Na I line
variations being caused by neutral matter funnelled onto the star
by magnetic fields. The variability of H$_\alpha$ and He I in
LkH$_\alpha$234 are due to episodic gas accretion, while the
observed variations in Na I lines is a transient event that
exhibits the dynamical signature of a body ($\geq$100 km in
diameter) falling onto the star. It is worth mentioning here that
recent studies on meteors from cometary origin by
Trigo-Rodriguez et al. (2004) have shown greater sodium abundances
than those expected for inter-planetary dust particles and
chondritic meteorites and Potter et al. (2002) have detected comet
like sodium tail from planet Mercury, and also the discovery of
presence of sodium tail in comets (Cremonese et al. 1997). Thus an
infalling solid body of asteroid size and deficient in H can
produce the observed Na I RAC and BEC and no correlation with the
H and He I lines.

The dusty circumstellar disk of the young star LkH$_\alpha$234
could be a newly formed protoplanetary disk consisting of solid
bodies of different sizes (Fuente et al. 2001). Theoretical models
of formation of planetary bodies in circumstellar disks around a
1M$_\odot$ mass star show that planetesimals of sizes of up to a
few hundred kilometers can quickly form within the first 10$^5$
years and the typical cumulative number of such bodies can be
10$^4$ to 10$^6$ between 0.5 to 1.5 AU from the star (Lissauer
1993, Weidenschilling et al. 1997). While such numbers are not
known for B type stars, LkH$_\alpha$234 (age $\sim$ 0.1 Myrs) may
have high frequency of infall events (Grady et al. 2000).

\section{Conclusion} We have presented here high resolution
spectroscopic monitoring of a Herbig Be star LkH$_\alpha$234 in
the Na I (D2\&D1), He I (5876\AA) and H$_\alpha$ lines. We found
no correlations between the Na I and the He I and H$_\alpha$ line
variability in the five data sets over a period of 31 days (7th
Oct. 2003 to 8th Nov. 2003) though we do find a correlation
between He I and H$_\alpha$. Thus the origin of variation in the
Na I line (RAC up to 200 km/s and BEC up to 100 km/s seen only on
13th Oct. 2003) is different from those of H$_\alpha$ and He I.
While the H$_\alpha$ and He I line variations are due to episodic
gaseous accretion, the Na I RAC indicate a dramatic transient
event of solid body infall on to the star and the BEC as the blown
away parts of the solid body not projected against the stellar
surface. Considering the Keplerian dynamics and the harsh
environment of a B5 star, we estimate that a solid body of size
$\geq$100 km broke up and disintegrated at a distance between 0.1
to 2.0 AU from the star. This makes LkH$_\alpha$234 the youngest
system ($\sim$0.1 Myrs) with evidence for protoplanetary bodies of
asteroidal size. We plan to pursue further spectroscopic
monitoring of the star with the HET to determine the frequency of
such events and further understand the complicated dynamics of
this very young circumstellar environment.



\acknowledgments The authors would like to thank Larry Ramsey for
additional discretionary time on the Hobby-Eberly Telescope and
Eric Feigelson and Mike Eracleous for important discussions on the
subject. The authors would also like to thank the referee Mike
Sitko for very valuable suggestions which made the content of the
paper strong. The Hobby-Eberly Telescope (HET) is a joint project
of the University of Texas at Austin, the Pennsylvania State
University, Stanford University, Ludwig-Maximilians-Universitat
Munchen, and Georg-August-Universitat Gottingen. The HET is named
in honor of its principal benefactors, William P. Hobby and Robert
E. Eberly. A.C. and J.G. would like to acknowledge National
Science Foundation for grant AST-01-38235, AST-02-43090 and NASA
for grants NAG5-12115 and NAG5-11427. S.M. would like to
acknowledge the support of the Michelson Fellowship Program (JPL,
NASA).

\clearpage


\begin{figure}
\epsscale{0.8} \plotone{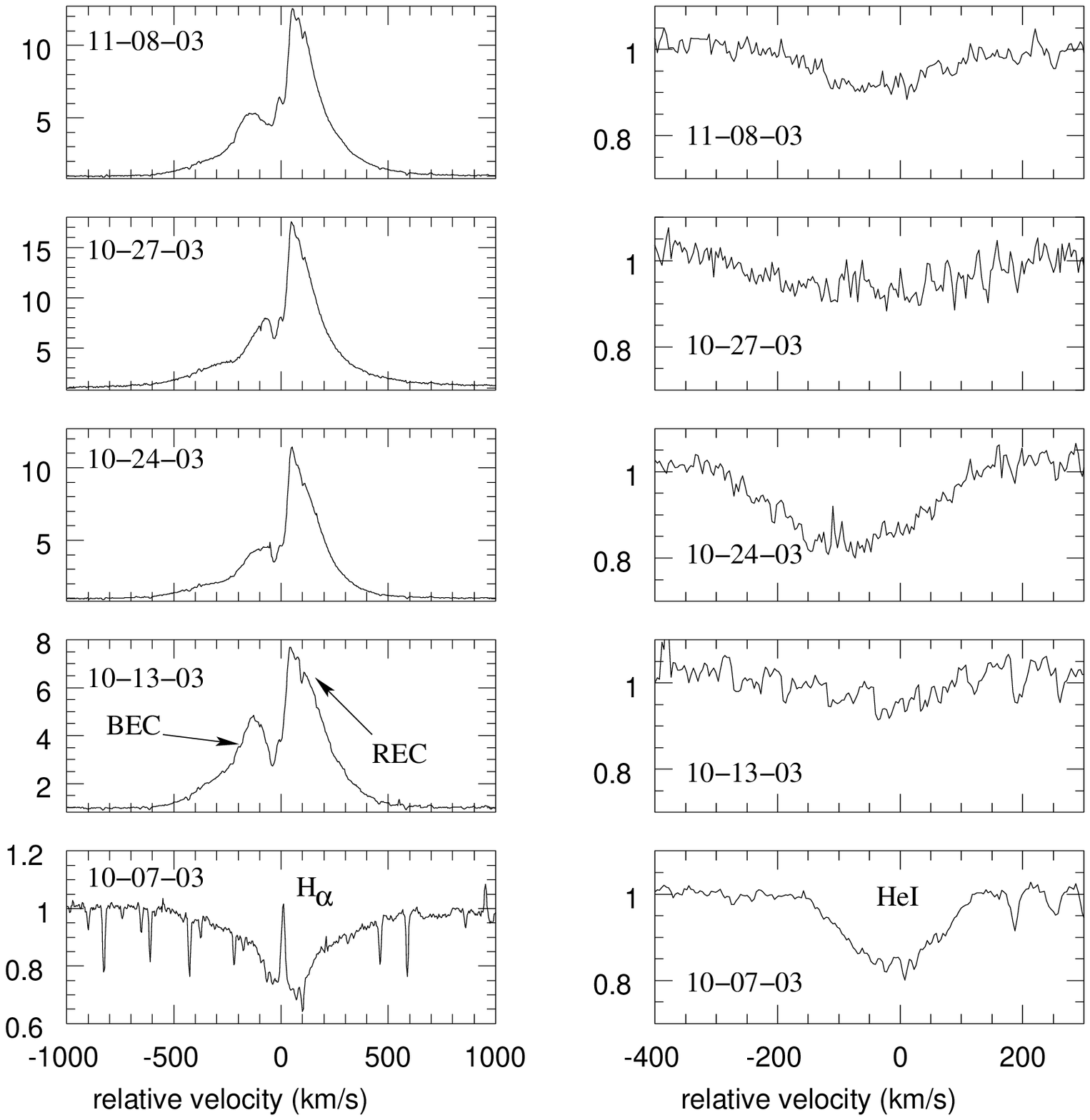} \caption{The left panel shows the
spectra of LkH$_\alpha$234 around the H$_\alpha$(6563\AA) emission
line and the right panel around HeI (5876\AA). The x-axis is in
relative velocity with respect to the rest wavelengths of
H$_\alpha$ and HeI respectively. The y-axis is the flux normalised
with respect to the stellar continuum. The dates of observations
are shown with each spectrum. The narrow 20 km/s width nebular H$_\alpha$
emission line is seen near the core of the stellar photospheric
absorption line on 10-07-03 and so are many nebular absorption
lines. The nebular H$_\alpha$ line can also be seen on the blue
wing of the REC on the other dates of observations.\label{fig1}}
\end{figure}

\begin{figure}
\epsscale{0.8} \plotone{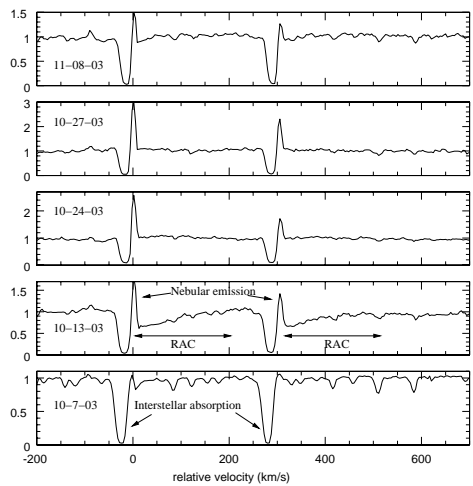} \caption{The spectra of
LkH$_\alpha$234 in the region of Na I D2(5890\AA) and D1(5896\AA)
absorption features. The x-axis has been converted into relative
velocity (km/s) with respect to the Na I D2 line at the stellar
rest frame. The y-axis is same as in figure 1. The Red shifted
Absorption Components (RAC), the nebular emission and the
interstellar absorption lines are marked. The dates of
observations are shown along with each spectrum. Also seen are the
very mild Blueshifted Absorption Components (BEC; on 10-13-03) on
the blue side of the Na I (D2\&D1) interstellar absorption lines.
The interstellar Na I lines are offset by 27 km/s from the rest
frame of the star.\label{fig2}}
\end{figure}

\clearpage




\end{document}